\documentclass{article}
\usepackage{spconf,amsmath,epsfig}
\usepackage{amssymb}
\usepackage{amsmath}
\usepackage{url}
\usepackage{xcolor, soul}
\sethlcolor{green}
\usepackage{tabularx}
\usepackage{multirow}
\usepackage{array}
\usepackage{bm}

\renewenvironment{thebibliography}[1]{%
   \begin{oldthebibliography}{#1}%
     \setlength{\itemsep}{-0.6ex}%
}%
{%
   \end{oldthebibliography}%
}

\begin{document}\sloppy

\def\x{{\mathbf x}}
\def\L{{\cal L}}
\newcommand{\etal}{\textit{et al}. }
\newcommand\T{\rule{0pt}{2.6ex}}       
\newcommand\B{\rule[-1.2ex]{0pt}{0pt}} 
\title{PHASE-INCORPORATING SPEECH ENHANCEMENT BASED ON COMPLEX-VALUED GAUSSIAN PROCESS LATENT VARIABLE MODEL}
%
\name{Sih-Huei Chen, Yuan-Shan Lee, and Jia-Ching Wang}
\address{}

\maketitle

\begin{abstract}
Traditional speech enhancement techniques modify the magnitude of a speech in time-frequency domain, and use the phase of a noisy speech to resynthesize a time domain speech. This work proposes a complex-valued Gaussian process latent variable model (CGPLVM) to enhance directly the complex-valued noisy spectrum, modifying not only the magnitude but also the phase. The main idea that underlies the developed method is the modeling of short-time Fourier transform (STFT) coefficients across the time frames of a speech as a proper complex Gaussian process (GP) with noise added. The proposed method is based on projecting the spectrum into a low-dimensional subspace. The likelihood criterion is used to optimize the hyperparameters of the model. Experiments were carried out on the CHTTL database, which contains the digits zero to nine in Mandarin. Several standard measures are used to demonstrate that the proposed method outperforms baseline methods.
\end{abstract}
\section{Introduction}
\label{sec:intro}

Speech enhancement is an important topic in the field of speech processing, and its purpose is to increase the quality and intelligibility of a noisy speech. Two major methods of representing a signal in the time-frequency (T-F) domain are used. The first is a statistical model-based method, which does not require prior knowledge about speech or noise signals, so has a low computational complexity. The second is a template-based method, in which the patterns of speech (noise) are stored in the pre-trained speech (noise) model. In these methods, T-F masking is commonly used to extract the speech component from a noisy signal. However, the masked signals still contain some noise. Some distortions of the speech occur because musical noise is generated by the T-F masking.

To solve these problems caused by T-F masking, various approaches have been developed for enhancing masked spectra. One of a well-known method is template-based method  \cite{Williamson_2013, Williamson_2014, Williamson_2014j, Wang_2016}. Williamson \etal \cite{Williamson_2013} utilized a binary mask to first separate the speech from background noise. Then, they employed a sparse representation (SR) method, which assumes that a magnitude spectrum of a speech is a linear combination of magnitude spectra that are formed by clean speech signals, to enhance the masked speech signal. Williamson \etal  \cite{Williamson_2014} generated a soft mask using a deep neural network (DNN). Then, they used non-negative matrix factorization (NMF), which represents the magnitude of a speech signal as a linear sum of a basis matrix and an activation matrix, to modify the masked speech signal. Recently, Wang \etal \cite{Wang_2016} presented a compressive sensing (CS)-based speech enhancement method. They used formant detection to obtain a binary mask. They then utilized an over-complete dictionary based on CS to impute the value associated with the missing frequency bin. Notably, all of the above mentioned template-based methods, utilized in the reconstruction stage, are applied only to the magnitude of the masked STFT coefficients, while phase is ignored. Additionally, they all consider a linear relationship between the speech spectrum and the corresponding weight which associated with speech components. However, recent investigations have demonstrated that taking into account the phase improves the quality of enhanced speech \cite{Gerkmann_2015}. Besides, linear model may not capture the nonlinear property of speech.

This work develops a two-stage method for speech enhancement. In the first stage, a binary mask is estimated using power spectral density (PSD). The masked complex-valued STFT coefficients of a speech signal are regarded as an incomplete spectra. In the second stage, a complex-valued Gaussian process latent variable model (CGPLVM) is proposed to reconstruct the incomplete spectra in a complex domain. The major contributions of this work are summarized as follows. 
\begin{itemize}
\item The speech spectra across time frames are modeled as a proper complex Gaussian process (GP), which provides a nonlinear mapping from a latent space which associated with speech components to speech space in the T-F domain. 
\item Rather than estimating the phase, the complex-valued STFT coefficients are directly estimated frame by frame that modifies both the magnitude and the phase of a noisy speech.
\end{itemize}

The remainder of the paper is organized as follows. Section \ref{sec:background} provides a mathematical description of the issue of interest and related studies are discussed. Section \ref{sec:gplvm speech enhancenment} describes the proposed two-stage speech enhancement that is based on GPLVM \cite{Lawrence_2006} and proposed CGPLVM. Section \ref{sec:Experimental} evaluates the performance of the proposed method using the CHTTL database. Finally, Section \ref{sec:Conclusions} draws conclusions and discusses future work.

\section{Background}
\label{sec:background}

Template-based speech enhancement methods \cite{Gonzalez_2014, Luo_2016, Min_2016} tend to process a signal in the T-F domain. A time-domain noisy signal $x(n)\in \mathbb{R}$ can be modeled as a clean signal $s(n)\in \mathbb{R}$ that is contaminated by a noise signal $n(n)\in \mathbb{R}, n \in \mathbb{Z}^+$ in the STFT domain, as follows. 
\begin{equation} \label{eq:noisymodel}
X(f,t)=S(f,t)+N(f,t)
\end{equation}
where $f$ and $t$ are the indices of the frequency bin and the frame, respectively. Eq. (\ref{eq:noisymodel}) can be rewritten as the product of a magnitude component and a phase component,
\begin{equation}
\left| {X(f,t)} \right|e^{j\varphi _X (f,t)} = \left| {S(f,t)} \right|e^{j\varphi _S (f,t)} + \left| {N(f,t)} \right|e^{j\varphi _N (f,t)}
\end{equation}
where $j = \sqrt { - 1}$, $\left| {\cdot} \right|$ denotes the magnitude and $\varphi _{\cdot} $ denotes the phase angle. In speech enhancement and robust speech recognition, T-F masking is a powerful way to reduce the effects of noise \cite{Gemmeke_2010, Josifovski_1999}. Two commonly used masks are the binary mask and the soft mask. A binary mask is defined as follows.
\begin{equation} \label{eq:mask}
M(f,t)=\left\{\begin{matrix}
1, &  {\rm{if}} \quad \dfrac{{\left| {\widehat S(f,t)} \right|}}{{\left| {\widehat S(f,t)} \right| + \left| {\widehat N(f,t)} \right|}} \ge c\\ 
0, & {\rm{otherwise}}
\end{matrix}\right.
\end{equation}
where $c$ is an empirically determined threshold. A larger $c$ results in the domination of more frequency bins by noise. $\widehat S(f,t)$ and $\widehat N(f,t)$ denote the estimated speech and noise, respectively. A soft mask is generated from a ratio of estimated speech magnitude to noisy speech magnitude, resulting in smooth masked spectra.

Let ${\mathbf{\widetilde S}} =  {\mathbf{M}}  \otimes  {\left|  \mathbf{X} \right|} \in {\mathbb{R}}^{F \times T}$ denote a masked spectrogram. NMF-based methods \cite{Williamson_2014, Gemmeke_etal_2011} generally assume that a spectrogram of speech can be reconstructed using a pre-trained basis matrix and a corresponding activation matrix. The activation matrix is obtained as follows.
\begin{equation}
{\mathbf{{\widehat H}}} =\textup{arg} \underset{\mathbf{H}}{\text{min}}\left \| \mathbf{\widetilde S}-\mathbf{WH} \right \|_F
\textup{ }\textup{ }\textup{ } s.t.\textup{ } {\mathbf{W}},{\rm{ }}{\mathbf{H}} \ge 0
\end{equation}
where ${\mathbf{W}}\in {\mathbb{R}}^{F \times K}$ is the pre-trained basis matrix; ${\mathbf{\widehat H}}\in {\mathbb{R}}^{K \times T}$ is the activation matrix, and $K$ is the number of basis vectors.

Similarly, the masked spectrogram can be enhanced by utilizing sparse representation (SR). SR \cite{Williamson_2013} imposes a different constraint on the activation matrix. An overcomplete dictionary $\mathbf{D}$ is typically used to reconstruct the speech signal. The sparse activation matrix $\mathbf{\widehat A}$ is given by
\begin{equation}
{\mathbf{\widehat A}}=\textup{arg} \underset{\mathbf{A}}{\text{min}}\left \| \mathbf{\widetilde S}-\mathbf{DA} \right \|_F
\textup{ }\textup{ }\textup{ } s.t. \textup{ } \forall i,\left \| \mathbf{a}_i \right \|_0\leq L
\end{equation}
where $\mathbf{a}_i$ is the $i$-th column vector of $\mathbf{A}$ and $L$ is a constant that controls sparseness.

After the estimated activation matrix has been obtained, the magnitude spectra of an instance of speech can be approximated as ${\dot{\mathbf{S}}} = \mathbf{W}\widehat{\mathbf{H}}$ (${\dot{\mathbf{S}}} = \mathbf{D}\widehat{\mathbf{A}}$). To resynthesize the time-domain signal, the phase information must be recovered. In various works \cite{Williamson_2013, Williamson_2014, Wang_2016}, the STFT coefficient of a speech signal is approximated as
\begin{equation}
S(f,t)\approx {\breve{S}(f,t)} = \dot{S}(f,t)e^{j\varphi _X (f,t)}
\end{equation}
Notably, $\varphi _X (f,t)$ represents the phase of the noisy signal. However, recent work has established that the resynthesized signal is inconsistent \cite{Gerkmann_2015}, meaning that $\rm{STFT}(iSTFT(\breve{\mathbf{S}}))\neq \breve{\mathbf{S}}$. 

The literature includes many template-based methods for dealing with the problem of inconsistency, which involves phase estimation \cite{Kameoka_2009, Magron_2016, Rodriguez_2016}. Kameoka \etal \cite{Kameoka_2009} proposed a complex NMF, which assumes that a complex-valued STFT coefficient is the product of two non-negative parameters with a phase term. An iterative algorithm has been developed to estimate phase. Magron \etal \cite{Magron_2016} considered a phase constraint in the framework of complex NMF to improve on the performance that was achieved by Kameoka \etal \cite{Kameoka_2009}. In summary, two points about existing methods are worthy of note: (1) the magnitude and phase are estimated separately in the real domain, and (2) only a linear model is considered. In this work, the feasibility and applicability of nonlinear model, named GPLVM, is first investigated for reconstructing the magnitude spectra. We then extend GPLVM to reconstruct directly complex-valued STFT coefficients that contain both magnitude and phase information.

\section{Gaussian process latent variable model (GPLVM)-based speech enhancement}
\label{sec:gplvm speech enhancenment}
Based on the work of Wang \etal \cite{Wang_2016}, this work presents a two-stage method for enhancing a noisy signal, which comprises a statistical model-based binary mask \cite{Martin_2001} and a nonlinear model for storing the pattern of speech. The following subsections will describe these steps in detail. 
\subsection{Missing data masks}
\begin{figure}[t]
\begin{minipage}[b]{0.48\linewidth}
  \centering
\centerline{\epsfig{figure=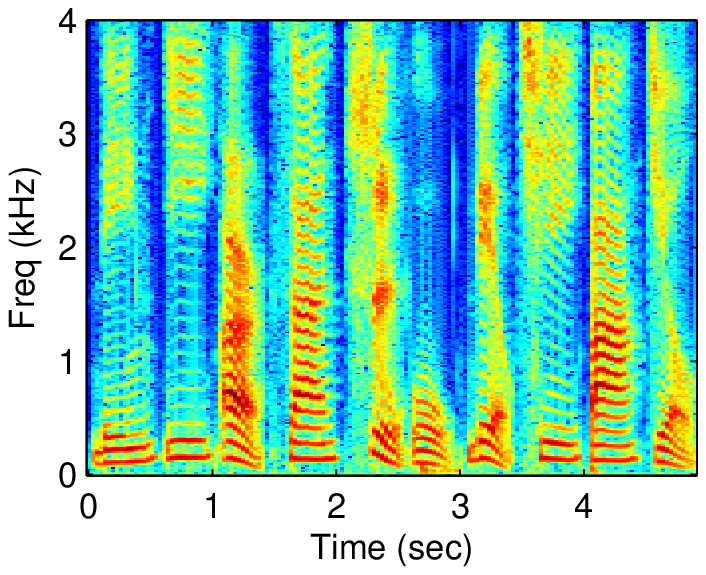,width=4.5cm,height=3.5cm}}
  \vspace{-0.1cm}
  \centerline{(a) }\medskip
\end{minipage}
\begin{minipage}[b]{0.48\linewidth}
  \centering
\centerline{\epsfig{figure=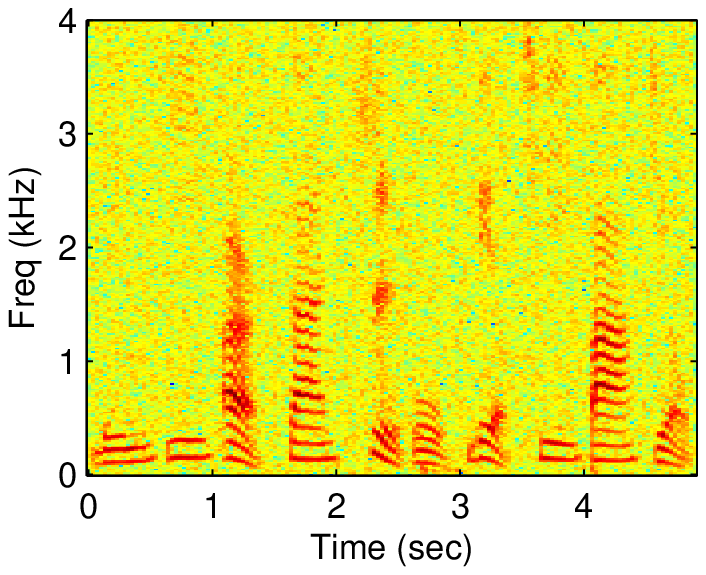,width=4.5cm,height=3.5cm}}
  \vspace{-0.1cm}
  \centerline{(b) }\medskip
\end{minipage}

\begin{minipage}[b]{0.48\linewidth}
  \centering
\centerline{\epsfig{figure=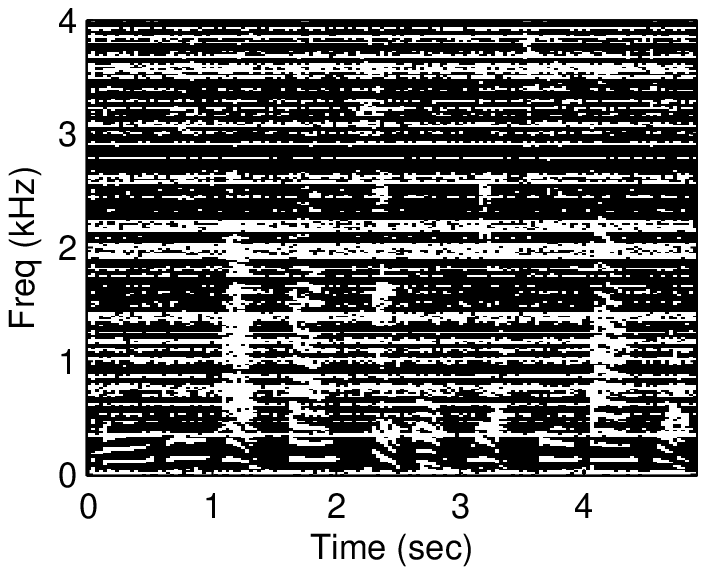,width=4.5cm,height=3.5cm}}
  \vspace{-0.1cm}
  \centerline{(c) }\medskip
\end{minipage}
\begin{minipage}[b]{0.48\linewidth}
  \centering
\centerline{\epsfig{figure=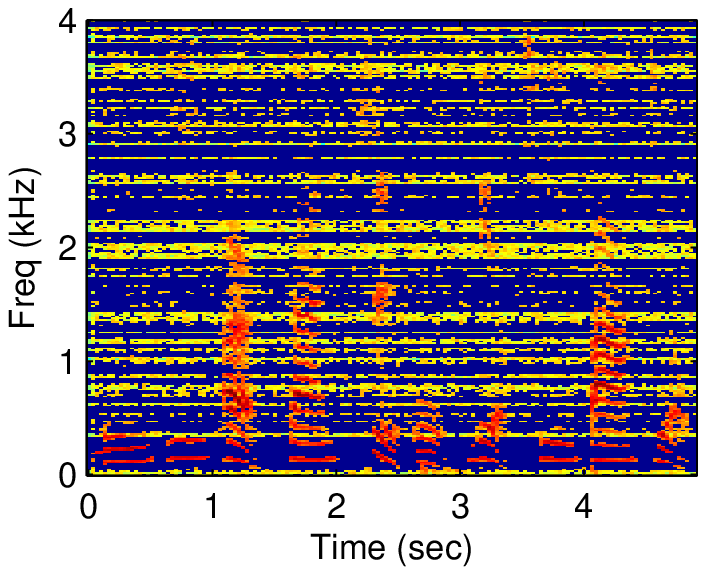,width=4.5cm,height=3.5cm}}
  \vspace{-0.1cm}
  \centerline{(d) }\medskip
\end{minipage}
\caption{Examples of (a) clean speech, (b) noisy speech, (c) missing data mask, and (d) incomplete observation.}
\label{fig:binarmaskedcleannoisy}
\end{figure}

Unlike in previous works \cite{Williamson_2013, Williamson_2014}, in which prior knowledge about the instance of speech and noise is used to generate a mask, in this work, a binary mask is estimated without training. Noise power spectral density (PSD) is utilized to determine whether an STFT bin is reliable or not. As in Eq. (\ref{eq:mask}), an element of the mask $\mathbf{M}$ is $1$, meaning that the corresponding SFTF bin is reliable. Otherwise, the STFT bin is unreliable. The masked spectra ${\mathbf{\widetilde S}}$ are then regarded as incomplete observations. Figure \ref{fig:binarmaskedcleannoisy} displays an example of how speech is estimated using a missing data mask. As in a previous work \cite{Wang_2016}, the goal here is to reconstruct speech spectra from the incomplete observations.

\subsection{GPLVM-based reconstruction of STFT magnitude}
First, we investigate the feasibility and applicability of nonlinear probabilistic model, named GPLVM \cite{Lawrence_2006}, for speech enhancement. In this subsection, the reconstruction is performed on magnitude spectrum. Each frequency band is independently regarded as a GP. GPLVM \cite{Lawrence_2006} is utilized to learn the T-F pattern of clean speech. Given training frames $\mathbf{Y}=[\mathbf{y}_1,...,\mathbf{y}_{QT}] \in {\mathbb{R}^{F \times QT}}$ which comprise $Q$ clean speech spectrograms, each frequency band $\mathbf{Y}_f\in {\mathbb{R}^{QT}}$ can be modelled as 
\begin{equation}
\mathbf{Y}_f=g_f(\mathbf{Z})+\bm{\epsilon}_f 
\end{equation}
where $\mathbf{Z}=[\mathbf{z}_1,...,\mathbf{z}_{QT}] \in \mathbb{R}^{K \times QT}$ with $K \ll F$, $\bm{\epsilon}_f {\sim} {\mathcal{N}}(\mathbf{0}, {\beta ^{ - 1}}{\mathbf{I}})$. The mappings $g_f, f = 1,...,F$ are drawn from an independent and identically distributed (i.i.d.) GP, i.e. $g_f(\mathbf{Z}) {\sim} {\mathcal{N}}(\mathbf{0}, \mathbf{K})$, where $\mathbf{K}$ is a covariance matrix in which the element of the $n$-th row and the $m$-th column is determined by a kernel function, $[\mathbf{K}]_{nm}=k(\mathbf{z}_n,\mathbf{z}_m)$,  $n, m \in \left \{1,...,QT  \right \}$. The form of a nonlinear mapping depends on the choice of a kernel function. For example, a radial basis function (RBF) kernel $k({\mathbf{z}}_n ,{\mathbf{z}}_m ) = \theta _1 \exp ( - \theta _2 \left\| {{\mathbf{z}}_n  - {\mathbf{z}}_m } \right\|^2 )$, where $\theta  = \{ \theta _1 ,{\rm{ }}\theta _2 \} $ are hyperparameters in the model, yields a smooth mapping. The marginal likelihood of $\mathbf{Y}$ can be calculated as
\begin{equation} \label{eq:marginalReal}
\begin{aligned}
 p(\left. {{\mathbf{Y}} } \right|{\mathbf{Z}}) = & {\prod_{f=1}^{F}} \int {p(\left. {{\mathbf{Y}}_f } \right|{\mathbf{g}}_f )} p(\left. {{\mathbf{g}}_f } \right|{\mathbf{Z}})d{\mathbf{g}}_f  \\ 
 = & {\prod_{f=1}^{F}} \mathcal{N}(\left. {{\mathbf{Y}}_f } \right|{\bf{0}},{\mathbf{K}}+{\beta}^{-1}\mathbf{I}) \\
\end{aligned} 
\end{equation}
where ${\mathbf{g}}_f = g_f(\mathbf{Z}) \in {\mathbb{R}^{QT}}$. The hyperparameters $\theta$ and the low-dimensional representations $\mathbf{Z}$ can be estimated by maximizing Eq. (\ref{eq:marginalReal}) by the gradient descend method \cite{Lawrence_2006}. Accordingly, the spectral patterns of the clean speech are then stored in the kernel.

Given estimation of $\theta$ and $\mathbf{Z}$, incomplete observations $\mathbf{\widetilde S}$ can be enhanced by calculating the corresponding low-dimensional representation. To reconstruct the speech spectrogram, the standard GP prediction is utilized \cite{Lawrence_2005recons}. The reconstructed spectrogram is then combined with the noisy phase.
\subsection{Phase-incorporating reconstruction of complex-valued STFT coefficient}
To incorporate the estimation of phase into the reconstruction, rather than modifying only the magnitude spectra, the complex-valued STFT coefficients are directly enhanced from the masked spectra ${\mathbf{\bar S}} =  {\mathbf{M}}  \otimes {\mathbf{X}}  \in {\mathbb{C}}^{F \times T}$. Let ${\mathbf{U}}=[\mathbf{u}_1,...,\mathbf{u}_{QT}]^\top \in {\mathbb{C}^{F \times QT}}$ be the complex-valued STFT coefficients of $Q$ training data from clean speech signals. Similar to GPLVM, each frequency band ${\mathbf{U}}_f$ can be viewed as a complex GP. To learn the nonlinear mapping between the complex-valued spectrum and its low-dimensional representation, this work proposes the CGPLVM. 
\begin{equation}
\mathbf{U}_f=h_f(\mathbf{V})+\mathbf{e}_f 
\end{equation}
where ${\mathbf{V}}=[\mathbf{v}_1,...,\mathbf{v}_{QT}]^\top \in {\mathbb{C}^{K \times QT}}$, $\mathbf{e}_f$ has a complex Gaussian distribution ${\mathcal{CN}}(\mathbf{0}, {\beta ^{ - 1}}{\mathbf{I}}, \mathbf{0})$ and $h_f, f=1,...,F$ are drawn from an i.i.d. proper complex GP, so $h_f(\mathbf{V}) {\sim} {\mathcal{CN}}(\mathbf{0}, \mathbf{K}_c, \mathbf{0})$. $\mathbf{K}_c$ is a kernel matrix that expresses the relationships among the complex-valued low-dimensional frames $\mathbf{v}_1,...,\mathbf{v}_{QT}$. Based on the work of Boloix-Tortosa \etal \cite{Boloix_2015}, a kernel that is used in a complex GP framework can be defined as
\begin{equation}
\begin{aligned}
 k_c ({\mathbf{v}}_n ,{\mathbf{v}}_m ) = & k_{rr} ({\mathbf{v}}_n ,{\mathbf{v}}_m ) + k_{jj} ({\mathbf{v}}_n ,{\mathbf{v}}_m ) \\
 & + j(k_{rj} ({\mathbf{v}}_m ,{\mathbf{v}}_n ) 
  - k_{rj} ({\mathbf{v}}_n ,{\mathbf{v}}_m ))  \\
\end{aligned}
\end{equation}
where $k_{rr}$, $k_{jj}$ and $k_{rj}$ are real kernel functions. In this work, a kernel that is the sum of an exponentiated quadratic kernel and a bias term is used.
\begin{equation}
k ({\mathbf{v}}_n ,{\mathbf{v}}_m ) = \theta _1\exp ( -\frac{({\mathbf{v}}_n  - {\mathbf{v}}_m )^\mathrm{H}({\mathbf{v}}_n  - {\mathbf{v}}_m )}{\theta _2})+\theta _3
\end{equation}
where $\left ( \cdot  \right )^\mathrm{H}$ is the Hermitian matrix. The hyperparameters and the low-dimensional representations can be learned by maximizing the log marginal likelihood as in Eq. (\ref{eq:marginalReal}). 
\begin{equation}
\begin{aligned}
 \ln p(\left. {\mathbf{U}} \right|{\mathbf{V}})
 & = - FQT\ln \pi  - F\ln \left| {{\mathbf{K}}_c+\beta^{-1}\mathbf{I} } \right| \\
 & - {\rm{trace}}(({{\mathbf{K}}_c+\beta^{-1}\mathbf{I} })^{ - 1} {\mathbf{UU}}^\mathrm{H} ) 
\end{aligned}
\end{equation}

By introducing the low-dimensional representations $\mathbf{V}$ that are associated with the training spectra $\mathbf{U}$ and the $t$-th frame $\mathbf{\bar s}_t$ from the masked spectra ${\mathbf{\bar S}}$, the corresponding low-dimensional representation $\mathbf{\bar v}_t$ can be obtained by solving the equation,
\begin{equation}
\widehat{\mathbf{v}}_t=\underset{\bar {\mathbf{v}}_t}{\arg \max }\ln p(\mathbf{U},\mathbf{\bar s}_t\mid \mathbf{V}, \mathbf{\bar v}_t)
\end{equation}
Finally, the $t$-th enhanced spectrum ${\breve{\mathbf{s}}_t}$ can be reconstructed using a predictive approach, which is given by ${\breve{\mathbf{s}}_t} = $ ${\mathbf{U}}^\mathrm{H}  ({{\mathbf{K}}_c+\beta^{-1}\mathbf{I} })^{ - 1} {\mathbf{k}}$, where ${\mathbf{k}} = [k_c ({\mathbf{v}}_1 ,{\widehat{\mathbf{v}}_t}),k_c ({\mathbf{v}}_2 ,{\widehat{\mathbf{v}}_t}),...,k_c ({\mathbf{v}}_{QT} ,{\widehat{\mathbf{v}}_t})]^\mathrm{T}$.

\section{Experimental Results}
\label{sec:Experimental}

\subsection{Experimental settings and performance metrics}
In this work, the performances of the proposed methods when applied to the CHTTL database \cite{CHTTL}, were evaluated. The CHTTL database includes 100 speakers (50 males and 50 females) who said the numbers zero to nine consecutively in Mandarin only once. Each complete utterance lasted 5-6 seconds, and was sampled at 8 kHz. In the experiments herein, 60 speakers (30 males and 30 females) were randomly selected from the CHTTL database. Data from ten (five males and five females) of them were used as training data. White noise was added with SNRs of 5, 10, 15 and 20 dB to the utterances of the remaining 50 speakers.

The performance of the tested enhancement method was evaluated in terms of segmental SNR (SSNR) \cite{Loizou:2013}, as
\begin{equation}
{\rm{SSNR}}=\frac{1}{T}\sum_{t=1}^{T}P\left \{ 10\log _{10}\frac{\left \| \mathbf{s}_t \right \|^2}{\left \| \mathbf{s}_t-\breve{\mathbf{s}}_t \right \|^2} \right \}
\end{equation}
where $\mathbf{s}_t$ and $\breve{\mathbf{s}}_t$ represent the clean speech and the enhanced speech in the $t$-th frame. $P(x)=\min \left \{ \max (x,\rm{-10}),\rm{35} \right \}$ confines the SNR in each frame to a perceptually meaningful range. The perceptual evaluation of speech quality (PESQ) \cite{Rix_2001} was used to measure the quality of speech. 

The spectrograms were generated using a 512-point STFT with Hamming windows to transform the speech into the time-frequency domain ($F=$ 257). The windows were shifted relative to each other by one half of the window length to cause them to overlap. 

The performance of the proposed method was compared with the following baselines.
\begin{itemize}
\item SR: A method of Williamson \etal \cite{Williamson_2013} was selected as the first baseline.  The SR was operated on magnitude spectra and the noisy phase was used to resynthesize the estimated speech signal. Using the settings of \cite{Williamson_2013}, an overcomplete dictionary is formed by concatenating the spectra of clean speech with the sparsity set to 5. ($L=5$ in Eq. (5)). The maximum number of iterations was set to 50.
\item NMF: Another method of Williamson \etal \cite{Williamson_2014} was selected as the second baseline. As in SR, only the magnitude was modified. Using the settings of \cite{Williamson_2014}, the number of vectors in the basis matrix was 80 and the Euclidean distance was used to calculate the reconstruction error. The maximum number of iterations was set to 40.
\end{itemize}

Notably, the above methods \cite{Williamson_2013, Williamson_2014} use a DNN-based mask to extract speech components. To ensure a fair comparison, the statistical model-based mask was utilized to generate the masked spectra.

\begin{figure}[t]
\begin{minipage}[b]{0.48\linewidth}
  \centering
\centerline{\epsfig{figure=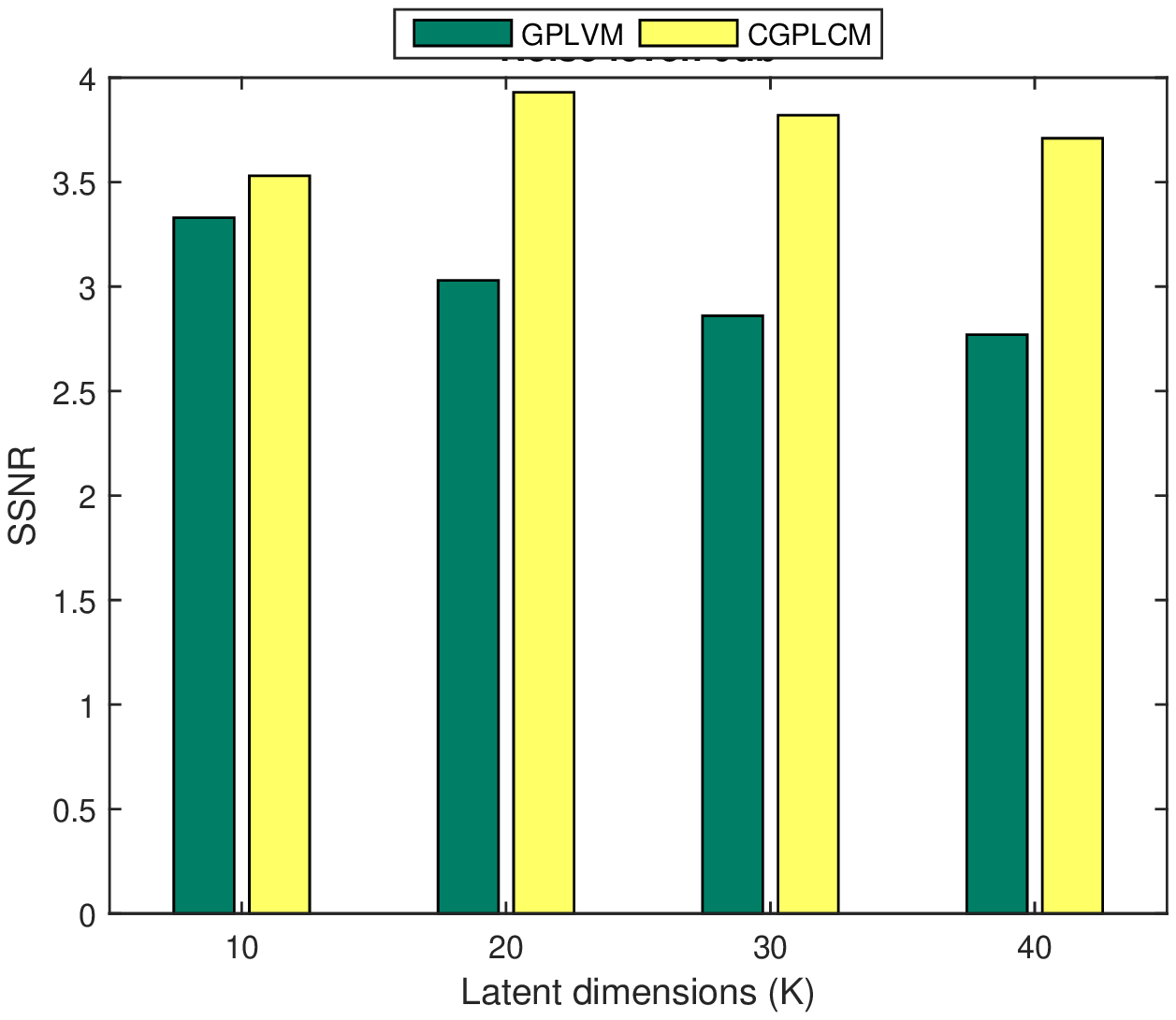,width=4.5cm,height=3.5cm}}
  \vspace{-0.1cm}
  \centerline{(a) }\medskip
\end{minipage}
\begin{minipage}[b]{0.48\linewidth}
  \centering
\centerline{\epsfig{figure=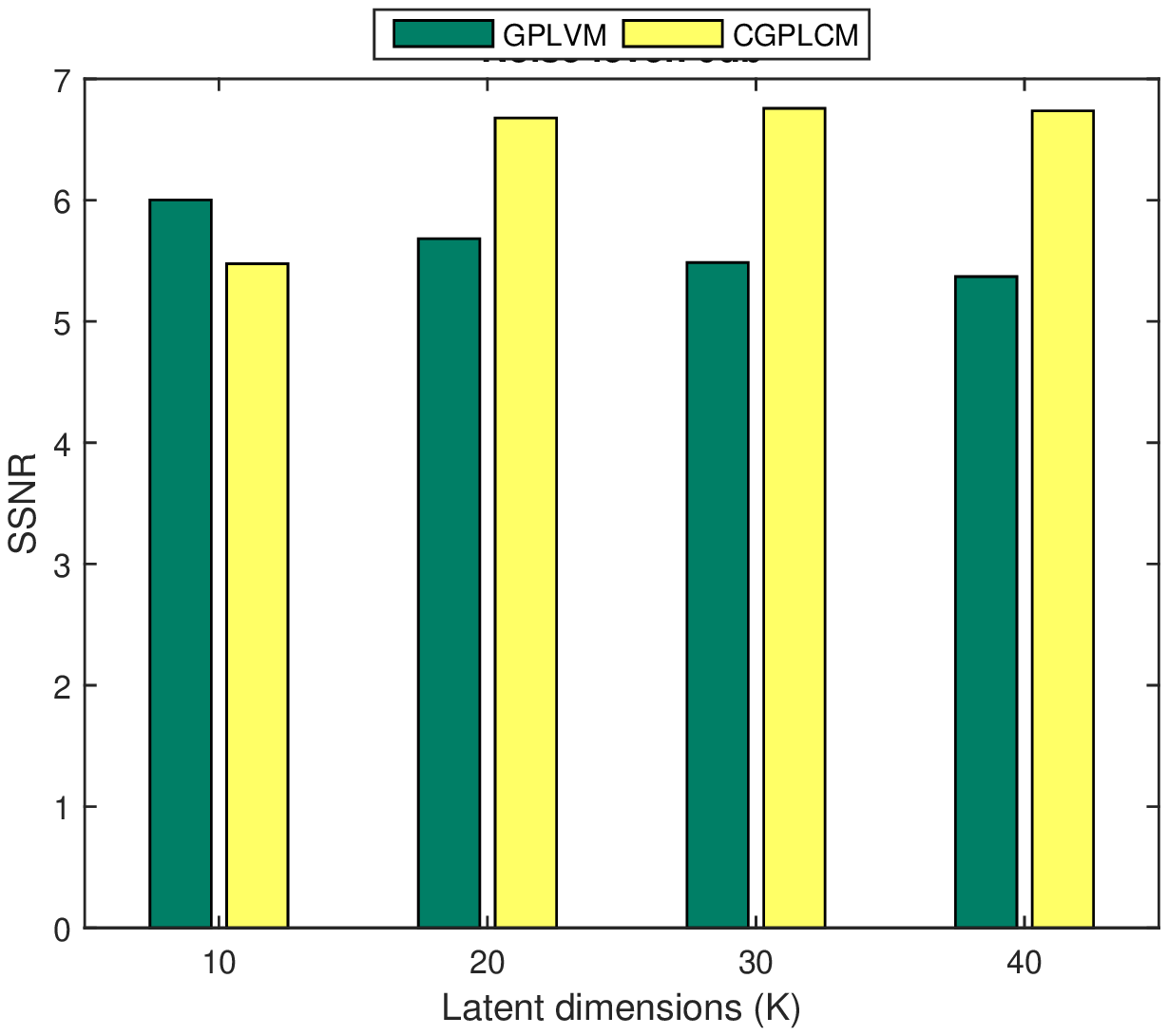,width=4.5cm,height=3.5cm}}
  \vspace{-0.1cm}
  \centerline{(b) }\medskip
\end{minipage}

\begin{minipage}[b]{0.48\linewidth}
  \centering
 \centerline{\epsfig{figure=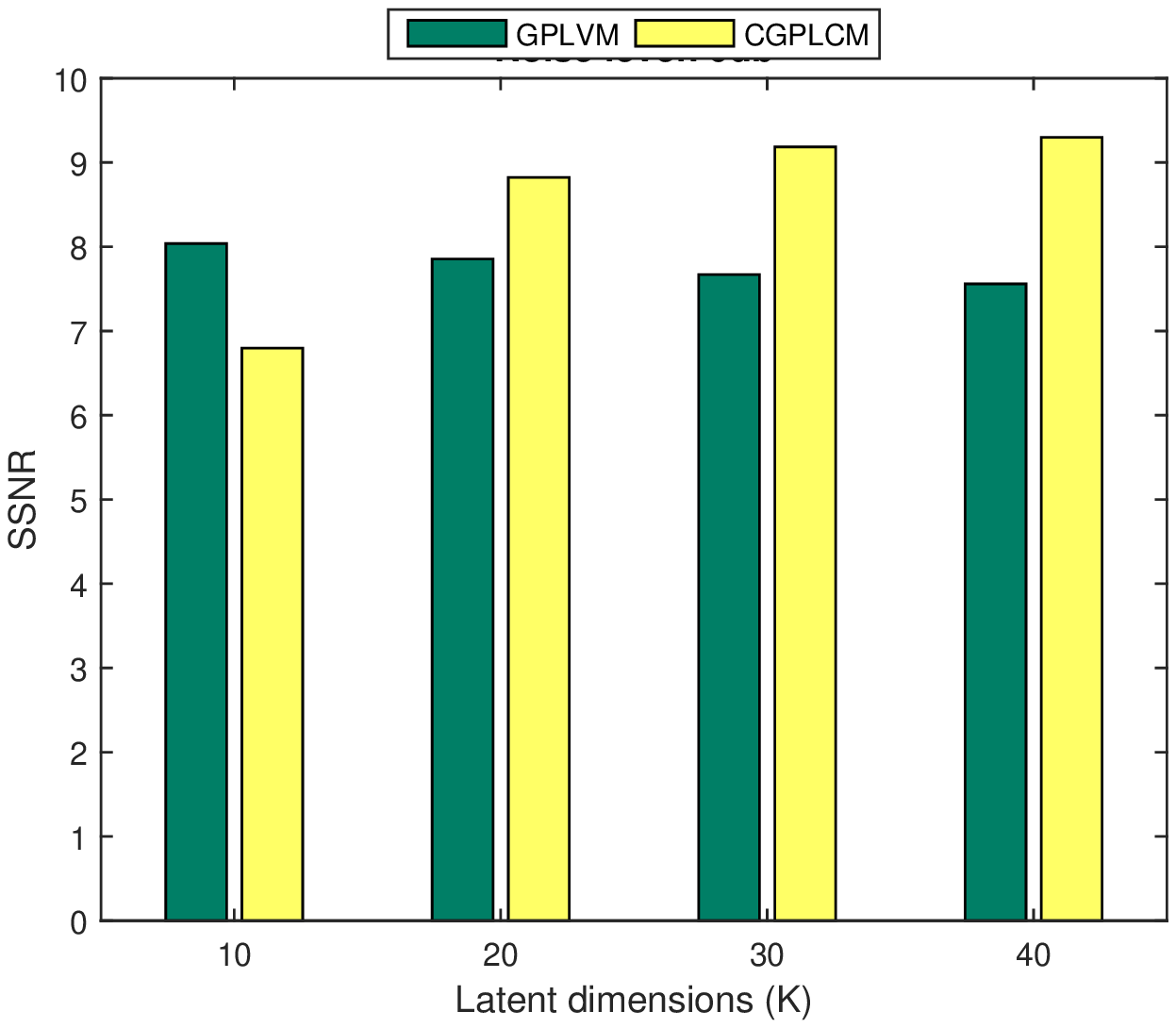,width=4.5cm,height=3.5cm}}
  \vspace{-0.1cm}
  \centerline{(c) }\medskip
\end{minipage}
\begin{minipage}[b]{0.48\linewidth}
  \centering
\centerline{\epsfig{figure=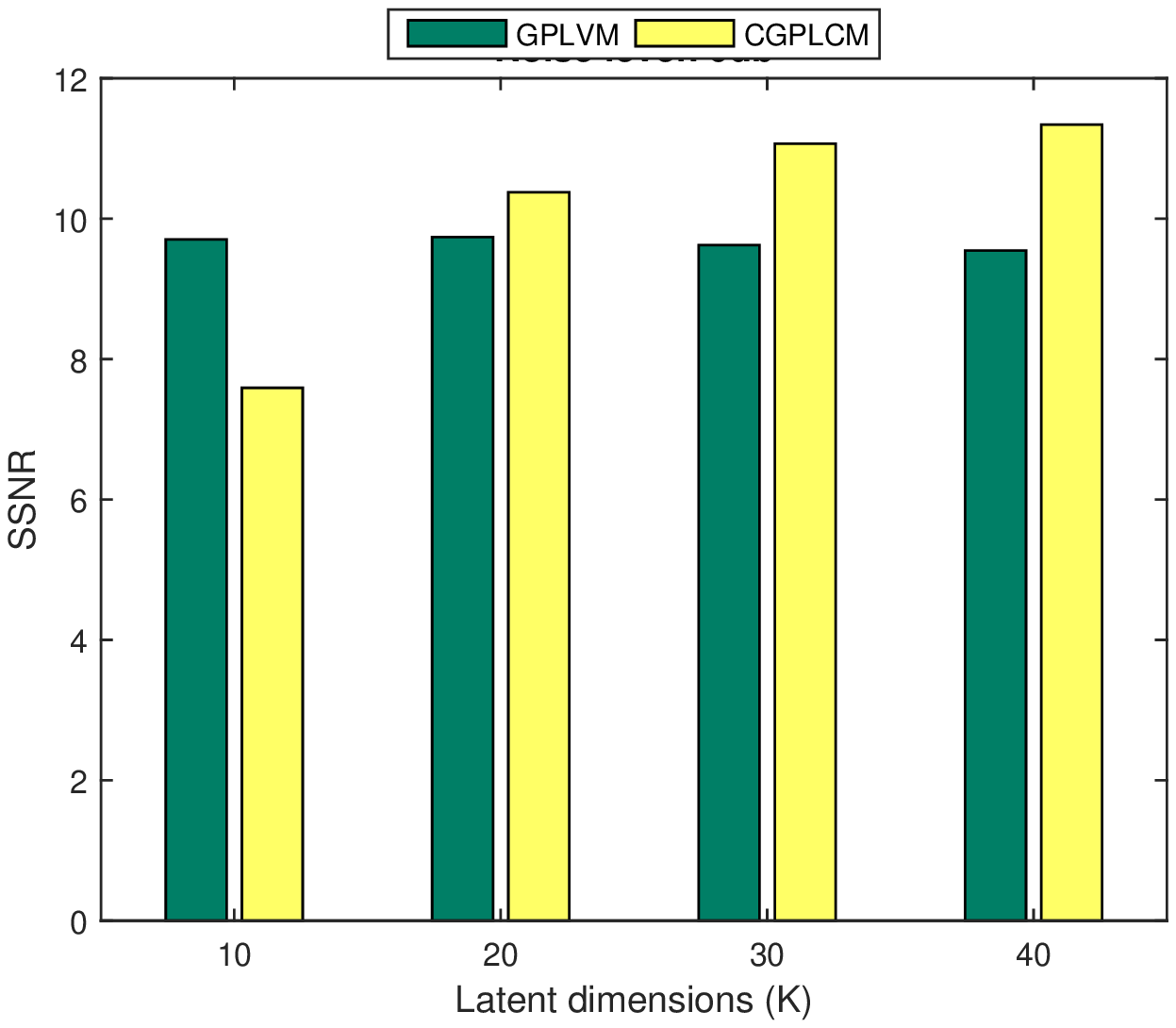,width=4.5cm,height=3.5cm}}
  \vspace{-0.1cm}
  \centerline{(d) }\medskip
\end{minipage}
\caption{Effects of latent dimension $K$ in proposed methods with various noise levels. (a) 5 dB , (b) 10 dB, (c) 15 dB, and (d) 20 dB.}
\label{fig:latentdimen}
\end{figure}

\subsection{Effects of latent dimension $K$}
In this experiment, the performance that is obtained using different values of the latent dimension ($K=$ 10, 20, 30, 40) is studied. The threshold was set to 0.95. Figure \ref{fig:latentdimen} shows the relevant experimental results. For the CGPLVM, with a high noise level (15 and 20 dB), a larger $K$ implies better performance and the highest SSNR is obtained at $K<40$ when the noise level is low (5 and 10 dB), perhaps because the speech components of the masked spectra are highly corrupted, reflecting the fact that the learned model with the highest latent dimension ($K=40$) was overfitting. Additionally, the GPLVM underperforms the CGPLVM except in the case of $K=10$.

\begin{figure}[t] 
\begin{minipage}[b]{0.48\linewidth}
  \centering
\centerline{\epsfig{figure=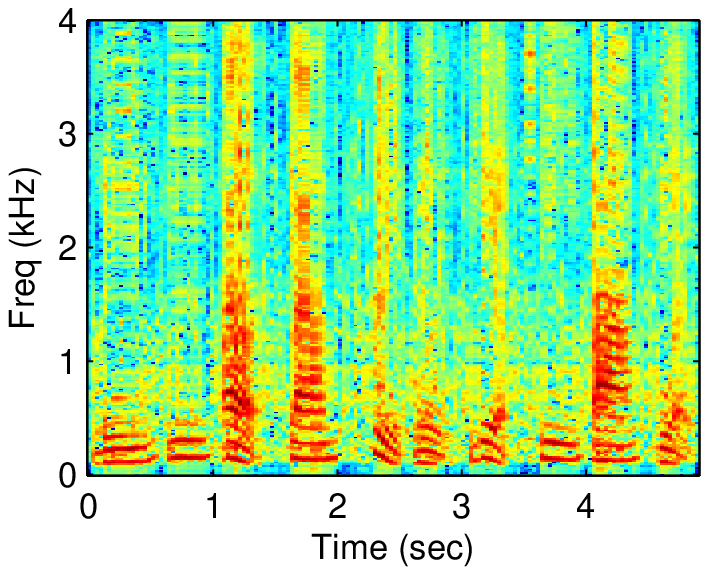,width=4.5cm,height=3.5cm}}
  \vspace{-0.1cm}
  \centerline{(a) }\medskip
\end{minipage}
\begin{minipage}[b]{0.48\linewidth}
  \centering
\centerline{\epsfig{figure=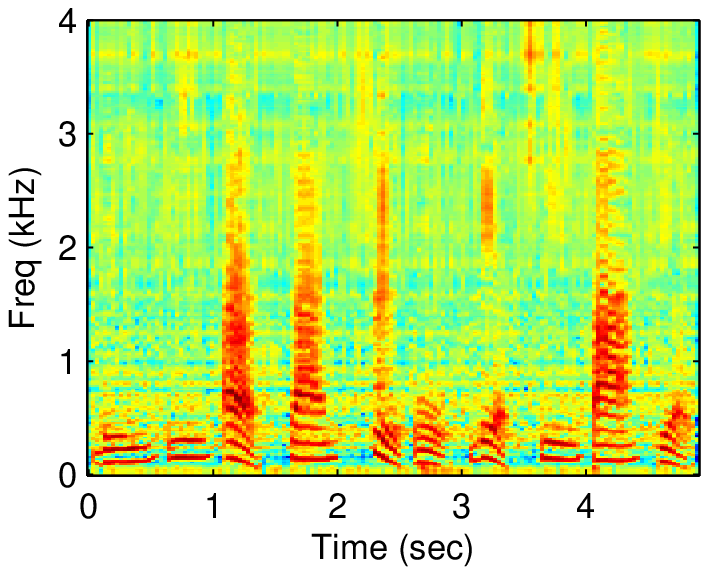,width=4.5cm,height=3.5cm}}
  \vspace{-0.1cm}
  \centerline{(b) }\medskip
\end{minipage}

\begin{minipage}[b]{0.48\linewidth}
  \centering
 \centerline{\epsfig{figure=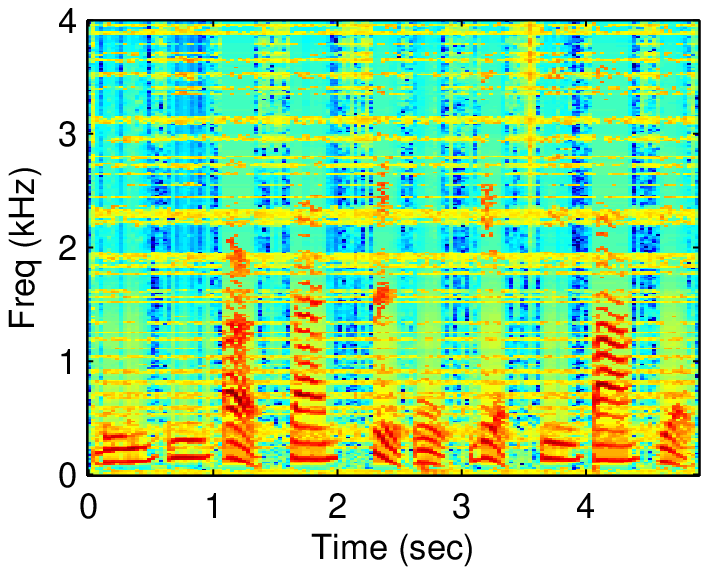,width=4.5cm,height=3.5cm}}
  \vspace{-0.1cm}
  \centerline{(c) }\medskip
\end{minipage}
\begin{minipage}[b]{0.48\linewidth}
  \centering
\centerline{\epsfig{figure=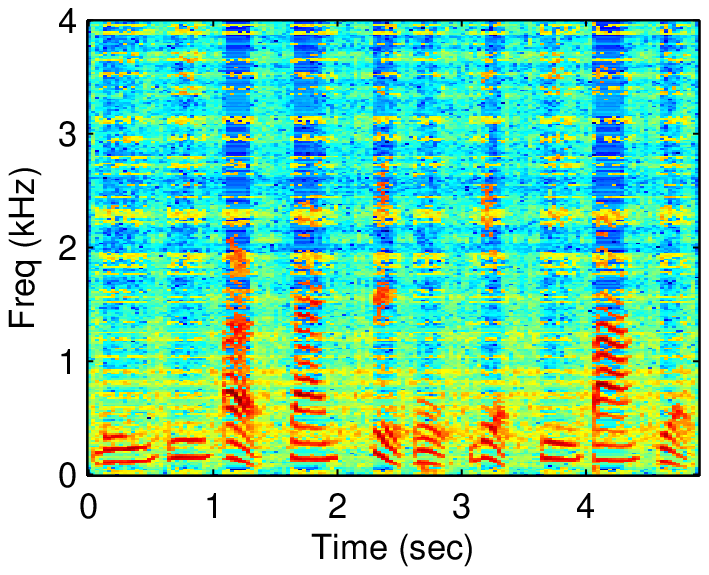,width=4.5cm,height=3.5cm}}
  \vspace{-0.1cm}
  \centerline{(d) }\medskip
\end{minipage}
\caption{Spectrogram reconstructed using (a) SR, (b) NMF, (c) GPLVM, and (d) CGPLVM. The noise level was set to 10 dB.}
\label{fig:ReconstructedSpectrogram}
\end{figure}

\begin{figure}[t]
\centering
\centerline{\epsfig{figure=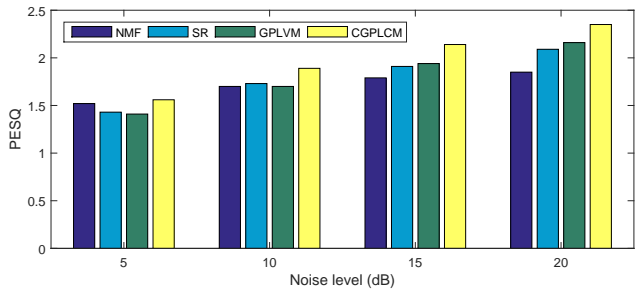,width=10cm}}
\caption{PESQs of proposed methods and baselines with various noise levels.}
\label{fig:PESQ}
\end{figure}

\subsection{Comparison of proposed methods and baselines}

For various values of threshold $c$, the proposed methods (GPLVM, CGPLVM) were compared with the baselines (SR, NMF) in terms of SSNR and PESQ. There are three values of the threshold ($c=$ 0.75, 0.85, and 0.95) were investigated. The latent dimension $K$ of the proposed methods was set to 30. Tables 1-3 present the experimental results. For all methods, the SSNR increases with the threshold perhaps because a higher threshold reflects the fact that the masked spectra contain less noise. In the other word, the masked spectra mainly consist of speech components that are consistent with the assumptions of the model. Experimental results reveal that the CGPLVM outperforms the baselines for various noise levels and thresholds.

\begin{table}[ht]
\caption{SSNR of proposed method and baselines with various noise level ($c=0.75$) } 
\centering 
\begin{tabular}{c@{\hskip 0.2in}c@{\hskip 0.2in}c@{\hskip 0.2in}c@{\hskip 0.2in}c@{\hskip 0.2in}c@{\hskip 0.2in}c} 
\hline\hline 
Noise level (dB) &  5            & 10           & 15           & 20           \T  \\ [0.5ex] 
\hline 
\centering SR \cite{Williamson_2013}                  & 2.22 & 3.81 & 5.23 & 6.35  \T \\ 
\centering NMF \cite{Williamson_2014}            & 1.34 & 3.62 & 6.07 & 7.99  \\ \hline
\centering GPLVM      & 1.04 & 3.71 & 6.49 & 9.01  \T \\ 
\centering CGPLVM    & \textbf{2.40} & \textbf{5.55} & \textbf{8.50} & \textbf{10.80}  \\ [0.5ex] 
\hline\hline 
\end{tabular}
\end{table}

\begin{table}[ht]
\caption{SSNR of proposed method and baselines with various noise level ($c=0.85$) } 
\centering 
\begin{tabular}{c@{\hskip 0.2in}c@{\hskip 0.2in}c@{\hskip 0.2in}c@{\hskip 0.2in}c@{\hskip 0.2in}c@{\hskip 0.2in}c} 
\hline\hline 
Noise level (dB) &  5            & 10           & 15           & 20           \T  \\ [0.5ex] 
\hline 
\centering SR \cite{Williamson_2013}                  & 2.83 & 4.14 & 5.34 & 6.38  \T \\ 
\centering NMF \cite{Williamson_2014}           & 2.33 & 4.13 & 6.06 & 8.24  \\ \hline
\centering GPLVM      & 1.80 & 4.16 & 6.68 & 9.08  \T \\ 
\centering CGPLVM    & \textbf{3.34} & \textbf{5.94} & \textbf{8.87} & \textbf{11.66}  \\ [0.5ex] 
\hline\hline 
\end{tabular}
\end{table}

\begin{table}[ht]
\caption{SSNR of proposed method and baselines with various noise level ($c=0.95$) } 
\centering 
\begin{tabular}{c@{\hskip 0.2in}c@{\hskip 0.2in}c@{\hskip 0.2in}c@{\hskip 0.2in}c@{\hskip 0.2in}c@{\hskip 0.2in}c} 
\hline\hline 
Noise level (dB) &  5            & 10           & 15           & 20           \T  \\ [0.5ex] 
\hline 
\centering SR \cite{Williamson_2013}                  & 3.28 & 4.93 & 5.93 & 6.69  \T \\ 
\centering NMF \cite{Williamson_2014}           & 3.17 & 5.36 & 7.11 & 8.64  \\ \hline
\centering GPLVM      & 2.86 & 5.49 & 7.67 & 9.62  \T \\ 
\centering CGPLVM    & \textbf{3.82} & \textbf{6.76} & \textbf{9.19} & \textbf{11.07}  \\ [0.5ex] 
\hline\hline 
\end{tabular}
\end{table}

Figure \ref{fig:ReconstructedSpectrogram} displays the reconstructed spectrogram using the proposed methods and baseline methods. Due to limitations of space, the following experiments are not conducted using all values of threshold. The threshold $c$ was set to 0.95. The latent dimension $K$ of the proposed methods was set to 30. For the CGPLVM, we take the magnitude of the reconstructed complex-valued STFT coefficients. Referring to the spectrogram of the clean speech (Fig. \ref{fig:binarmaskedcleannoisy} (a)), nonlinear model-based methods (GPLVM and CGPLVM) outperformed linear model-based methods (SR and NMF).

To demonstrate the superiority of the proposed methods, which jointly estimate the magnitude and phase of a speech, the PESQs that were obtained using the proposed methods and baselines were evaluated. The results in Figure \ref{fig:PESQ} demonstrate that the CGPLVM achieved a better PESQ than the other methods which do not consider the phase information of a speech at any noise level.

\section{Conclusions}
\label{sec:Conclusions}

This paper develops two latent variable model based methods for speech enhancement. The potential of using a nonlinear model and a phase-incorporating nonlinear model for reconstructing a masked speech was studied. Unlike state-of-the-art template-based methods, the proposed methods herein directly enhances the complex-valued STFT coefficients of a speech signal in the complex domain, rather than separately enhancing the magnitude and phase in the real domain. Additionally, instead of using a dictionary, the method uses a kernel matrix, which specifies a GP, to store the clean speech patterns that provide a nonlinear relationship between the speech and the corresponding low-dimensional representation. Experimental results indicate that the proposed methods have significantly higher SSNR and PESQ values than two baseline methods. In the future, we will consider other types of noise, such as babble, market, and piano. Besides, the model can be extended to deeper architectures to further boost its performance.

%

\bibliographystyle{IEEEbib}
\bibliography{icme2017template}

\begin{thebibliography}{10}

\bibitem{Williamson_2013}
D.~S. Williamson, Y.~Wang, and D.~Wang,
\newblock ``A sparse representation approach for perceptual quality improvement
  of separated speech,''
\newblock in {\em Proc. ICASSP}, 2013, pp. 7015--7019.

\bibitem{Williamson_2014}
D.~S. Williamson, Y.~Wang, and D.~Wang,
\newblock ``A two-stage approach for improving the perceptual quality of
  separated speech,''
\newblock in {\em Proc. ICASSP}, 2014, pp. 7034--7038.

\bibitem{Williamson_2014j}
D.~S. Williamson, Y.~Wang, and D.~L. Wang,
\newblock ``Reconstruction techniques for improving the perceptual quality of
  binary masked speech,''
\newblock {\em J. Acoust. Soc. Amer.}, vol. 136, pp. 892--902, 2014.

\bibitem{Wang_2016}
J.~C. Wang, Y.~S. Lee, C.~H. Lin, S.~F. Wang, C.~H. Shih, and C.~H. Wu,
\newblock ``Compressive sensing-based speech enhancement,''
\newblock {\em IEEE/ACM Trans. Audio, Speech, Language Process.}, vol. 24, no.
  11, pp. 2122--2131, 2016.

\bibitem{Gerkmann_2015}
T.~Gerkmann, M.~Krawczyk-Becker, and J.~Le Roux,
\newblock ``Phase processing for single-channel speech enhancement: History and
  recent advances,''
\newblock {\em IEEE Signal Process. Mag.}, vol. 32, no. 2, pp. 55--66, March
  2015.

\bibitem{Lawrence_2006}
Neil~D. Lawrence,
\newblock ``{The Gaussian process latent variable model},''
\newblock {\em Technical Report no CS-06-05}, 2006.

\bibitem{Gonzalez_2014}
S.~Gonzalez and M.~Brookes,
\newblock ``Mask-based enhancement for very low quality speech,''
\newblock in {\em Proc. ICASSP}, 2014, pp. 7029--7033.

\bibitem{Luo_2016}
Y.~Luo, G.~Bao, Y.~Xu, and Z.~Ye,
\newblock ``Supervised monaural speech enhancement using complementary joint
  sparse representations,''
\newblock {\em IEEE Signal Process. Lett.}, vol. 23, no. 2, pp. 237--241, 2016.

\bibitem{Min_2016}
G.~Min, X.~Zhang, J.~Yang, W.~Han, and X.~Zou,
\newblock ``A perceptually motivated approach via sparse and low-rank model for
  speech enhancement,''
\newblock in {\em Proc. ICME}, 2016, pp. 1--6.

\bibitem{Gemmeke_2010}
J.~F. Gemmeke, H.~Van Hamme, B.~Cranen, and L.~Boves,
\newblock ``Compressive sensing for missing data imputation in noise robust
  speech recognition,''
\newblock {\em IEEE Trans. Signal Process.}, vol. 4, no. 2, pp. 272--287, April
  2010.

\bibitem{Josifovski_1999}
L.~Josifovski, M.Cooke, P.~Green, and A.Vizinho,
\newblock ``State based imputation of missing data for robust speech
  recognition and speech enhancement,''
\newblock in {\em Proc. Eurospeech}, 1999, pp. 2837--2840.

\bibitem{Gemmeke_etal_2011}
J.~F. Gemmeke, T.~Virtanen, and A.~Hurmalainen,
\newblock ``Exemplar-based sparse representations for noise robust automatic
  speech recognition,''
\newblock {\em IEEE Trans. Audio, Speech, Language Process.}, vol. 19, no. 7,
  pp. 2067--2080, Sept 2011.

\bibitem{Kameoka_2009}
H.~Kameoka, Nobutaka Ono, Kunio Kashino, and Shigeki Sagayama,
\newblock ``Complex nmf: A new sparse representation for acoustic signals,''
\newblock in {\em Proc. ICASSP}, 2009, pp. 3437--3440.

\bibitem{Magron_2016}
P.~Magron, R.~Badeau, and B.~David,
\newblock ``Complex nmf under phase constraints based on signal modeling:
  Application to audio source separation,''
\newblock in {\em Proc. ICASSP}, 2016, pp. 46--50.

\bibitem{Rodriguez_2016}
F.~J. Rodriguez-Serrano, S.~Ewert, P.~Vera-Candeas, and M.~Sandler,
\newblock ``A score-informed shift-invariant extension of complex matrix
  factorization for improving the separation of overlapped partials in music
  recordings,''
\newblock in {\em Proc. ICASSP}, 2016, pp. 61--65.

\bibitem{Martin_2001}
R.~Martin,
\newblock ``Noise power spectral density estimation based on optimal smoothing
  and minimum statistics,''
\newblock {\em IEEE Trans. Speech, Audio Process.}, vol. 9, no. 5, pp.
  504--512, 2001.

\bibitem{Lawrence_2005recons}
Neil Lawrence,
\newblock ``Probabilistic non-linear principal component analysis with gaussian
  process latent variable models,''
\newblock {\em J. Mach. Learn. Res.}, vol. 6, pp. 1783--1816, 2005.

\bibitem{Boloix_2015}
R.~Boloix{-}Tortosa, F.~J. Payan{-}Somet, E.~Arias{-}de{-}Reyna, and
  J.~Jos{\'{e}} Murillo{-}Fuentes,
\newblock ``Proper complex gaussian processes for regression,''
\newblock {\em CoRR}, vol. abs/1502.04868, 2015.

\bibitem{CHTTL}
``{CHTTL} database,'' \url{http://www.aclclp.org.tw/use_mat_c.php#chttl}.

\bibitem{Loizou:2013}
Philipos~C. Loizou,
\newblock {\em Speech Enhancement: Theory and Practice},
\newblock CRC Press, Inc., Boca Raton, FL, USA, 2nd edition, 2013.

\bibitem{Rix_2001}
M.~P.~Hollier A.~W.~Rix, J. G.~Beerends and A.~P. Hekstra,
\newblock ``Perceptual evaluation of speech quality (pesq)-a new method for
  speech quality assessment of telephone networks and codecs,''
\newblock in {\em Proc. ICASSP}, 2001, pp. 749--752.

\end{thebibliography}

\end{document}